\documentclass[twocolumn,aps,superscriptaddress]{revtex4}
\usepackage{graphicx}
\usepackage{float}
\usepackage{dcolumn}
\usepackage{bm}
\usepackage{color}
\usepackage{amsmath}
\usepackage{amssymb}
\usepackage{amsfonts}
\usepackage{esint}
\usepackage{times}
\usepackage{xcolor}
\usepackage{phaistos}
\usepackage{braket}
\usepackage{comment}

\usepackage{pdfpages}

\usepackage[colorlinks,linkcolor=blue,anchorcolor=blue,citecolor=blue]{hyperref}

\begin{document}
\title{Suppression of static $ZZ$ interaction in an all-transmon quantum processor}
\author{Peng Zhao}\email{shangniguo@sina.com}
\affiliation{National Laboratory of Solid State Microstructures, School of Physics, Nanjing University, Nanjing 210093, China}
\author{Dong Lan}\email{land@nju.edu.cn}
\affiliation{National Laboratory of Solid State Microstructures, School of Physics, Nanjing University, Nanjing 210093, China}
\author{Peng Xu}
\affiliation{Institute of Quantum Information and Technology,
Nanjing University of Posts and Telecommunications, Nanjing, Jiangsu 210003, China}
\author{Guangming Xue}
\affiliation{Beijing Academy of Quantum Information Sciences, Beijing 100193, China}
\author{Mace Blank}
\affiliation{National Laboratory of Solid State Microstructures, School of Physics, Nanjing University, Nanjing 210093, China}
\author{Xinsheng Tan}\email{tanxs@nju.edu.cn}\affiliation{National Laboratory of Solid State Microstructures, School of Physics, Nanjing University, Nanjing 210093, China}
\author{Haifeng Yu}
\affiliation{Beijing Academy of Quantum Information Sciences, Beijing 100193, China}
\author{Yang Yu}
\affiliation{National Laboratory of Solid State Microstructures, School of Physics, Nanjing University, Nanjing 210093, China}

\date{\today}

\begin{abstract}

The superconducting transmon qubit is currently a leading qubit modality
for quantum computing, but gate performance in quantum processor with
transmons is often insufficient to support running complex algorithms for
practical applications. It is thus highly desirable to further improve
gate performance. Due to the weak anharmonicity of transmon, a static $ZZ$ interaction
between coupled transmons commonly exists, undermining the gate performance,
and in long term, it can become performance limiting. Here
we theoretically explore a promising parameter region in an
all-transmon system to address this issue. We show that an
feasible parameter region, where the $ZZ$ interaction is heavily suppressed
while leaving $XY$ interaction with an adequate strength to implement two-qubit
gates, can be found for all-transmon systems. Thus, two-qubit gates, such as
cross-resonance gate or iSWAP gate, can be realized without the detrimental
effect from static $ZZ$ interaction. To illustrate this, we
demonstrate that an iSWAP gate with fast gate speed and dramatically lower conditional
phase error can be achieved. Scaling up to large-scale transmon quantum processor,
especially the cases with fixed coupling, addressing error, idling error, and
crosstalk that arises from static $ZZ$ interaction could also be strongly suppressed.

\end{abstract}

\maketitle

\section{Introduction}\label{ch1}

The transmon qubit has been demonstrated as a leading superconducting qubit
modality for quantum computing since it has been largely responsible
for the recent impressive achievements in superconducting quantum
information processing \cite{R1,R2,R3}. These achievements crucially rely on the
improvement of the gate performance through increasing the transmon
coherence time \cite{R4} and mitigating coherent error from non-ideal
parasitic interaction \cite{R5,R6,R7,R8}. Nonetheless, state-of-the-art gate performance in
transmon quantum processor so far is probably insufficient to demonstrate
quantum advantage for practical applications \cite{R9} and to achieve the long-term goals
of fault-tolerant quantum computing \cite{R10}. This indicates that considerable effort
devoted to improving gate performance is still required.

Despite the benefit of the reproducibly long coherence times, the weak anharmonicity
of transmons currently poses a significant challenge to further improve
gate performance. For single qubit gates,
due to the weak anharmonicity, higher-energy levels of transmon can be easily populated
during microwave driven gate operation, causing leakage error. By using the
Derivative Removal by Adiabatic Gate (DRAG) scheme \cite{R11,R12},
this issue can be substantially mitigated without scarifying the gate speed, and
single qubit gate with fidelity above 99.9$\%$ can be achieved \cite{R1,R4}. However, various
pending challenges for pursuing two-qubit gates with fast speed and high fidelity
still exist due to non-ideal parasitic interactions that arise from the weak anharmonicity
of transmons \cite{R13,R14}. One of the leading non-ideal parasitic interaction is the
static $ZZ$ coupling \cite{R15,R16}, that has been shown to undermine
the performance of $XY$ interaction based two-qubit gates, such as cross-resonance gate or iSWAP
gate \cite{R2,R7,R8}. Meanwhile, its residual can cause idling error, and produce quantum
crosstalk related to neighboring spectator qubits \cite{R17,R18,R19,R20}.
Hopefully, these idling errors and crosstalk can be strongly suppressed by
using a tunable coupler \cite{R6,R21,R22,R23}. More Recently, it has been demonstrated, both
experimentally \cite{R24,R25,R26} and theoretically \cite{R27,R28,R29}, that this static $ZZ$ can also be
suppressed heavily by coupling superconducting qubits with opposite-sign of anharmonicity.
However, mitigating static $ZZ$ coupling for $XY$-based
two-qubit gate operations or qubit architecture with fixed inter-qubit coupling
is still an outstanding challenge for all-transmon quantum processors \cite{R28}.

In this work, we show that suppressing static $ZZ$ coupling for two-qubit
gate operations can be achieved by engineering quantum interference in
an all-transmon quantum processor. Contrary to the commonly accepted view that
for all-transmon quantum processors, static $ZZ$ interaction can vanish only
by turning off inter-qubit coupling \cite{R14,R28}. We find that an experimentally
accessible parameter region, where the static $ZZ$ coupling is heavily suppressed
while leaving $XY$ interaction with an adequate strength to implement two-qubit
gates can be found in an all-transmon system.

\section{System Hamiltonian}\label{ch2}

\begin{figure}[tbp]
\begin{center}
\includegraphics[width=8cm,height=2cm]{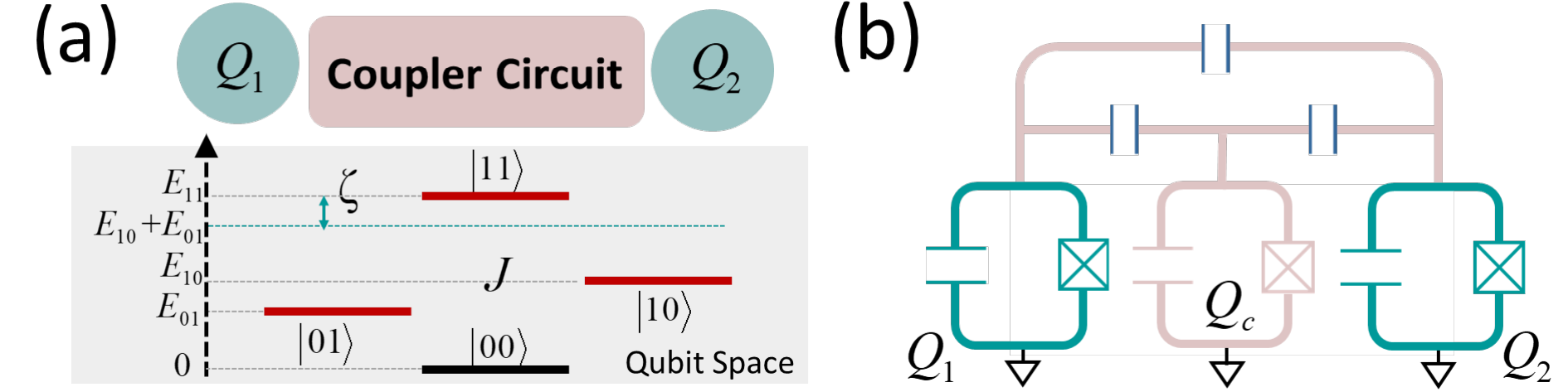}
\end{center}
\caption{(a) Circuit layout and level diagram of a coupled two-qubit system. Truncation to qubit
space gives rise an effective two-qubit Hamiltonian with an inter-qubit $XY$ coupling $J$ and a static $ZZ$
coupling $\zeta$. (b) Example circuit diagram of two transmons coupled via a coupler circuit combining
a capacitor and a ancilla transmon.}
\label{fig1}
\end{figure}

Since a superconducting qubit is naturally a multi-level system, especially for qubits
with weak anharmonicity such as the transmon, its higher-energy levels have a non-negligible
effect on qubit dynamics. For coupled transmons system shown in Fig.~\ref{fig1}(a),
truncation to the qubit (computational) subspace gives rise an effective two-qubit Hamiltonian with
not only an inter-qubit $XY$ coupling $J$ but also a static $ZZ$ coupling $\zeta$ that mainly arises from
interactions between qubit states and higher-energy states \cite{R16,R27}.
To suppress this $ZZ$ coupling for $XY$-based two qubit gates,
here we consider the all-transmon system schematically depicted in Fig.~\ref{fig1}(b),
where transmons $Q_{1(2)}$ are coupled through a coupling circuit combing a ancilla transmon $Q_{c}$
and a capacitor. The full system can be modeled by three coupled weakly anharmonic oscillators \cite{R30},
and described by (hereafter $\hbar =1$)
\begin{equation}
\begin{aligned}
\label{eq1}
H=&\sum_{j}\big(\tilde{\omega}_{j}q_{j}^{\dagger}q_{j}+\frac{\alpha_{j}}{2}q_{j}^{\dagger}q_{j}^{\dagger}q_{j}q_{j}\big)
\\&+\sum_{j\neq k}g_{jk}(q_{j}+q_{j}^{\dagger})(q_{k}+q_{k}^{\dagger}),
\end{aligned}
\end{equation}
where subscript $j(k)=\{1,2,c\}$ labels transmon $Q_{j}$ with anharmonicity
$\alpha_{j}$ and bare qubit frequency $\tilde{\omega}_{j}$, $q_{j}\,(q_{j}^{\dagger})$ is
the associated annihilation (creation) operator, and $g_{jk}$ denotes
strength of the coupling between $Q_{j}$ and $Q_{k}$.

\begin{figure}[tbp]
\begin{center}
\includegraphics[width=8cm,height=2.5cm]{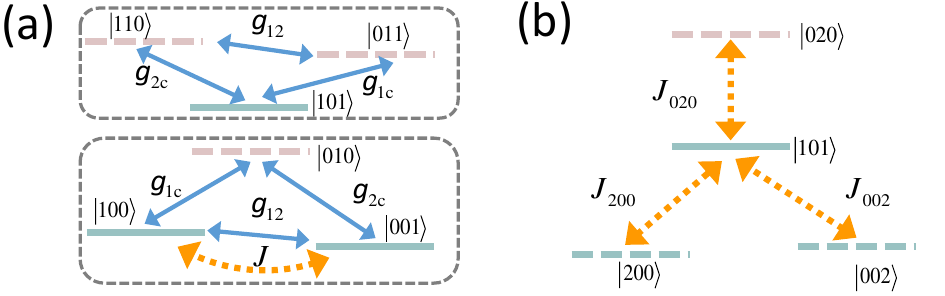}
\includegraphics[width=8cm,height=4cm]{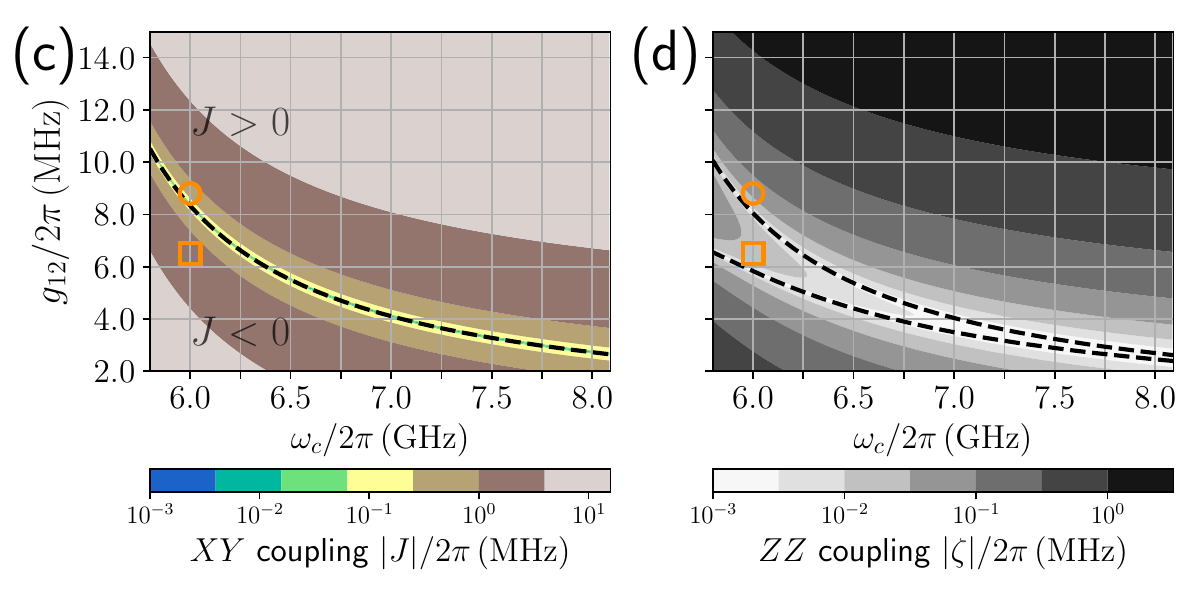}
\end{center}
\caption{(a)(b) Level diagram of the coupled transmon system shown in Fig.~\ref{fig1}(b). (c) Landscapes
of $XY$ coupling $J$ and (d) $ZZ$ coupling $\zeta$ (perturbation theory,
making RWA) as a function of coupler frequency $\omega_{c}$ and direct coupling strength $g_{12}$
with qubit frequency $\tilde{\omega}_{1(2)}/2\pi=5.114\,(4.914)\,\rm{GHz}$,
qubit/coupler anharmonicity $\alpha_{1(2)}/2\pi=-330\,\rm{MHz}$, $\alpha_{c}/2\pi=0\,\rm{MHz}$ (linear coupler),
and qubit-coupler interaction strength $g_{1c(2c)}/2\pi=98\,(83)\,\rm{MHz}$ \cite{R7,R35}. The dashed curves in (c) and (d) correspond to contours of $J=0$ and
$\zeta=0$, respectively. The open circle (square) in (c,d) mark the
set of parameters chosen to suppress $XY$ ($ZZ$) coupling.}
\label{fig2}
\end{figure}

We further consider that our system operates in the dispersive regime, i.e, the transom-coupler detuning $|\Delta_{1(2)}|=|\tilde{\omega}_{1(2)}-\tilde{\omega}_{c}|\gg g_{1c(2c)}$,
and the two transmons are in the straddling regime, i.e., transmon-tranmson detuning $|\Delta_{12}|=|\tilde{\omega}_{1}-\tilde{\omega}_{2}|<|\alpha_{1(2)}|$ \cite{R23,R31}.
Hence, truncation to two-qubit subspace, the Hamiltonian in Eq.~(\ref{eq1}) can be approximated by an
effective two-qubit Hamiltonian
\begin{eqnarray}
\begin{aligned}
\label{eq2}
H=\omega_{1}\frac{ZI}{2}+\omega_{2}\frac{IZ}{2}+J\frac{XX+YY}{2}+\zeta\frac{ZZ}{4},
\end{aligned}
\end{eqnarray}
where $(X,Y,Z,I)$ denote the Pauli operators and identity operator, and
the order indexes the qubit number, and $\omega_{1(2)}$ represents
the dressed qubit frequency of $Q_{1(2)}$. The last two terms
corresponds to the $XY$ coupling with strength $J$ and $ZZ$ coupling
with strength $\zeta$, respectively. As shown in Fig.~\ref{fig2}(a),
the $XY$ coupling results from the direct coupling
$g_{12}$ and the coupler-mediated indirect coupling, and its strength can be
approximated as $J=g_{12}+g_{1c}g_{2c}/\Delta$ with
$1/\Delta=(1/\Delta_{1}+1/\Delta_{2})/2$ \cite{R6,R22}. The $ZZ$ coupling comes from the interaction
between qubit states and non-qubit states (including higher-energy states of transmons and coupler
states), and its strength is defined as $\zeta=(E_{101}-E_{100})-(E_{001}-E_{000})$, where $E_{mnl}$
denotes the energy of system eigenstate $|mnl\rangle$ ($m,n,l=\{0,1,2\}$), and can be
perturbatively obtained \cite{R32,R33,R34}. Making the rotating-wave approximation (RWA),
and deriving up to the fourth-order perturbation gives an approximated expression $\zeta\simeq \zeta_{020}+\zeta_{200}+\zeta_{002}+\zeta_{1}$
(see Appendix A for details) with
\begin{eqnarray}
\begin{aligned}
\label{eq3}
&\zeta_{020}=\frac{J_{020}^{2}}{\Delta_{1}+\Delta_{2}-\alpha_{c}},\,
\zeta_{200}=\frac{J_{200}^{2}}{\Delta_{12}-\alpha_{2}},
\\&\zeta_{002}=-\frac{J_{002}^{2}}{\Delta_{12}+\alpha_{1}},\,
\zeta_{1}=\frac{4g_{12}g_{1c}g_{2c}}{\Delta_{1}\Delta_{2}},
\end{aligned}
\end{eqnarray}
where terms $\zeta_{020}$ and $\zeta_{002(200)}$ result from the effective coupling between qubit state $|101\rangle$
and higher-energy states of coupler $|020\rangle$ and transmons $|002(200)\rangle$, respectively, and
$J_{020},\,J_{200(002)}$ denote the associated effective coupling strength, as shown in Fig.~\ref{fig2}(b).

\section{suppression of static ZZ interaction for off-resonantly coupled Transmons}\label{ch3}

\begin{figure*}[tbp]
\begin{center}
\includegraphics[width=18cm,height=8cm]{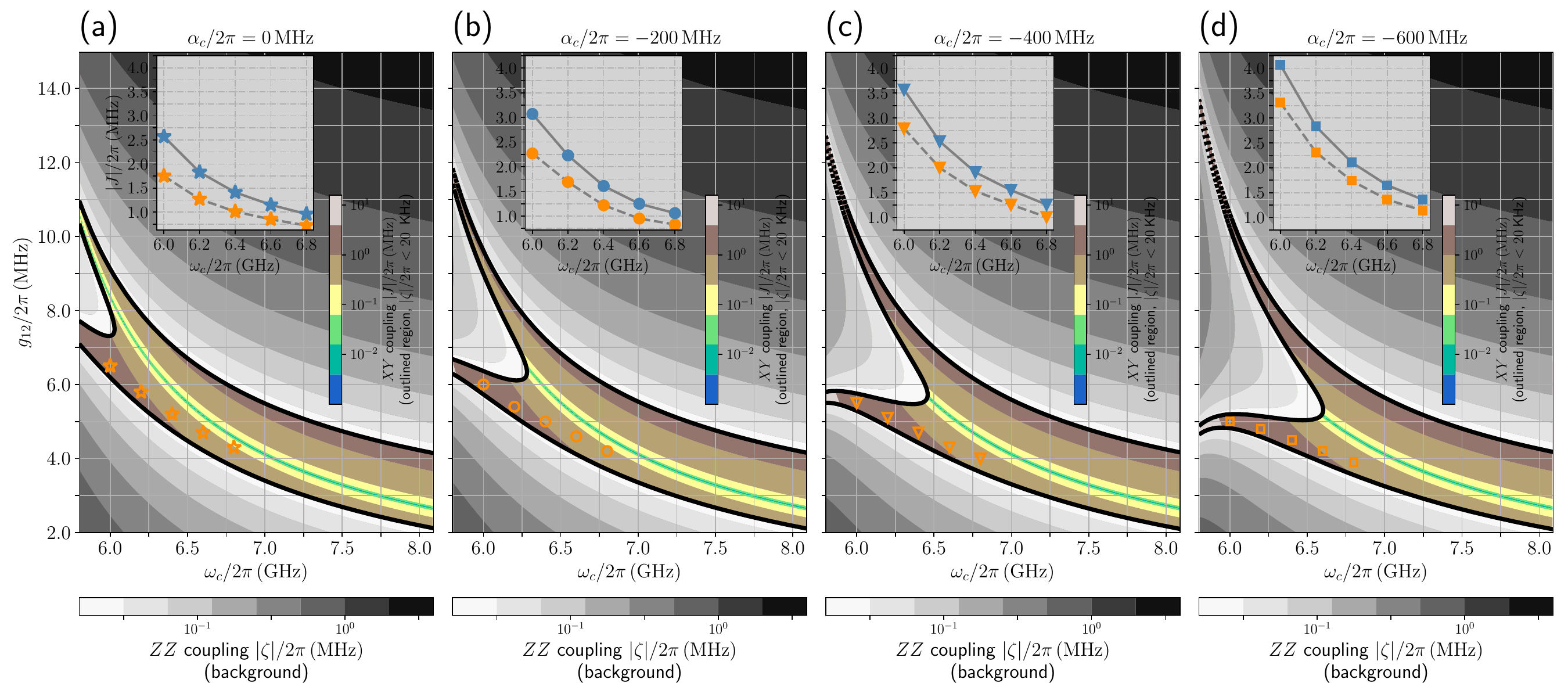}
\end{center}
\caption{ Landscapes of $XY$ coupling $J$ (perturbation theory, without making RWA) and $ZZ$ coupling $\zeta$ (numerical
diagonalization) as a function of coupler frequency $\omega_{c}$ and direct coupling strength $g_{12}$ with
coupler anharmonicity (a) $\alpha_{c}/2\pi=0$ (linear coupler), (b) $\alpha_{c}/2\pi=-200\,\rm{MHz}$,
(c) $\alpha_{c}/2\pi=-400\,\rm{MHz}$, and (d) $\alpha_{c}/2\pi=-600\,\rm{MHz}$. The other
system parameters are similar to those used in Fig.~\ref{fig2}. For each coupler anharmonicity, the two landscapes
are superposed in a single panel, where the region corresponding to ZZ coupling with $\zeta$ below 20 KHz is
omitted and replaced with the corresponding values of the XY coupling $J$ (the region outlined with black bold lines).
The Insets of (a-d) show
the maintained $XY$ coupling strength obtained with the corresponding parameter set
marked in (a-d). The solid (dashed) line represents the result derived from perturbation theory (the
period $T$ of the simulated cross-resonance oscillation). }
\label{fig3}
\end{figure*}

First, we consider that transmons are off-resonantly coupled together. According to perturbation
analysis \cite{R32,R33,R34}, Figures~\ref{fig2}(c) and~\ref{fig2}(d) show $J$ and $\zeta$ as a function of coupler frequency
$\omega_{c}$ and direct coupling strength $g_{12}$ with $\tilde{\omega}_{1(2)}/2\pi=5.114\,(4.914)\,\rm{GHz}$,
$\alpha_{1(2)}/2\pi=-330\,\rm{MHz}$, $\alpha_{c}/2\pi=0\,\rm{MHz}$ (linear coupler),
and $g_{1c(2c)}/2\pi=98\,(83)\,\rm{MHz}$ \cite{R7,R35}.
One can find that in the parameter space of ($\omega_{c},\,g_{12}$), zero-point for $XY$ coupling
forms a single branch (dashed line in Fig.~\ref{fig2}(c)), and when the coupler has
larger transition frequency, the zero-$ZZ$ regime almost overlap with the zero-$XY$ branch,
conforming the commonly accepted view, i.e., static $ZZ$ interaction can be suppressed by
turning off inter-qubit coupling. However, interestingly and unexpectedly, with decreasing the coupler
frequency, the zero-$ZZ$ regime gradually diverging and eventually splitting into two separate branches
(dashed lines in Fig.~\ref{fig2}(d)), and the presence of the lower branch shows that $ZZ$ coupling can be
heavily suppressed without the need for suppressing $XY$ coupling, thus one can mitigate static $ZZ$ coupling
for implementing $XY$-based two-qubit gates.

The physics behind the above interesting features are: (i) As shown in Eq.~(\ref{eq3}), when
coupler has larger transition frequency, the energy of coupler state $|020\rangle$ is far
lager than that of the states $|101\rangle$ and $|200(002)\rangle$, i.e, $E_{020}\gg \{E_{101},E_{200},E_{002}\}$, and
$\Delta_{1(2)}\gg \{\Delta_{12},\alpha_{1(2)}\}$, the terms $\zeta_{020}$ and $\zeta_{1}$
can be omitted. Thus, the dominated contribution to the total $ZZ$ coupling results from the effective
interaction between qubit state $|101\rangle$ and higher-energy states $|200\rangle$ ($|002\rangle$),
whose strength can be approximated as $J_{200(002)}\sim \sqrt{2}J$ (see Appendix A for details). When the inter-qubit $XY$
coupling $J$ is tuned off, the interactions between $|101\rangle$ and $|200\rangle$ ($|002\rangle$)
are also turned off effectively, causing suppression of $ZZ$ coupling.

(ii) However, when decreasing the coupler frequency, thus $E_{020}\gtrsim \{E_{101},E_{200},E_{002}\}$
and $\Delta_{1(2)}\gtrsim \{\Delta_{12},\alpha_{1(2)}\}$, the terms $\zeta_{020}$ and $\zeta_{1}$
can no longer be neglected. Since $E_{020}> E_{101} >\{E_{200},E_{002}\}$, the interaction
$|101\rangle\leftrightarrow|200\rangle(|002\rangle)$ gives rise an $ZZ$ coupling term with a positive
sign, while the interaction $|101\rangle\leftrightarrow|020\rangle$ contributes
a term with a negative sign. Therefore, the conditions for suppressing
$ZZ$ coupling cannot be achieved by just turning off the $XY$ coupling. To the opposite, the $XY$
coupling should be tuned on, thus the positive contribution and the negative contribution can interfere
destructively, giving rise the suppression of net $ZZ$ coupling. Moreover, since
$\zeta_{200(002)}\propto(\sqrt{2}J)^{2}$ (see Appendix A for details), for a fixed coupler frequency, there should be two
different values of $g_{12}$ (giving rise a total $XY$ coupling of same magnitude but opposite signs)
for suppressing $ZZ$ coupling. Up to now, the discussion above suggests that
the zero-$ZZ$ regime should be split into two separate branches which are approximately
symmetrical with respect to the zero-$XY$ branch. Thus for a given coupler frequency,
the strength of the maintained $XY$ coupling in the lower and upper branch of the zero-$ZZ$ region
should be approximately equal. However, as shown in Eq.~(\ref{eq3}), along with the terms
associated with higher-energy state of transmons and coupler, there is an addition terms
$\zeta_{1}$ associated with lower-energy states giving rise a positive contribution
with strength $\zeta_{1}\propto g_{12}$ (see Eq.~(\ref{eq3})). Thus, taking all these terms in Eq.~(\ref{eq3}) into consideration,
the strength of the maintained $XY$ coupling in the lower branch should be larger than that
in the upper branch, as shown in Fig.~\ref{fig2}(d).

To numerically verify the above results, we consider two sets of
parameters, i.e., $(6.0,6.5)$ and $(6.0,8.8)$, which are chosen to
suppress $XY$ and $ZZ$ coupling, respectively.
The $ZZ$ coupling strength can be exactly obtained by numerical diagonalization of
the system Hamiltonian in Eq.~(\ref{eq1}), giving rise $|\zeta|/2\pi=3.9\, (3.3)\,\rm{KHz}$
for the two parameter sets. while for $XY$ coupling, we note
that the effective $XY$ coupling is not perfectly well-defined for present system with
off-resonantly coupled transmons. Here the $XY$ coupling strength is estimated from the
period $T$ of the simulated cross-resonance oscillation with
the controlled qubit $Q_{1}$ in its ground state (see Appendix B for details). In the weak-drive limit,
the period $T$ can be well approximated by $2\pi/T=J\Omega_{d}/\Delta_{12}$ \cite{R35}, where
$\Omega_{d}$ denotes strength of the driving applied on $Q_{1}$.
As such, the estimated $XY$ coupling strength are $J/2\pi=1.75\,(0.63)\,\rm{MHz}$.

The above numerical results confirm that for systems operating on the lower branch of
zero-$ZZ$ region, the $ZZ$ coupling can indeed be strongly suppressed without the need for
suppressing the $XY$ coupling heavily. However, the maintained $XY$ coupling (here is
$1.75\,\rm{MHz}$) is too weak to support implementing a successful two-qubit gate,
such as the cross-resonance gate \cite{R36}. To achieve a larger maintained $XY$ coupling
for a given coupler frequency, according to the above analysis, one can reduce the energy
of $|020\rangle$ by replacing the liner coupler ($\alpha_{c}=0$) with a nonlinear coupler
with $\alpha_{c}<0$, such as a transmon. Then the negative $ZZ$ contribution ($\zeta_{020}$)
from interaction $|101\rangle\leftrightarrow|020\rangle$
gets larger, and a larger $XY$ coupling $J$ is thus needed to suppress $ZZ$ coupling.
Figure~\ref{fig3} show the numerical calculated $ZZ$ coupling strength and the $XY$ coupling strength
as a function of $\omega_{c}$ and $g_{12}$ with coupler anharmonicity $\alpha_{c}/2\pi=\{0,\,-200,\,-400,\,-600\}\,\rm{MHz}$.
The other parameters are same to those used in Fig.~\ref{fig2}. In order to easily identify
the desired parameter region for suppressing $ZZ$ coupling, for each coupler anharmonicity,
the associated two landscapes ($XY$ and $ZZ$ where date point with $ZZ$ coupling
strength below $20\,\rm{KHz}$ is removed) are superposed in a single panel. One can find
that the lower branch of zero-$ZZ$ region is shifted downward along with the increased
coupler anharmonicity. Thus, the maintained $XY$ coupling is also increased, as
shown in the inset of Fig.~\ref{eq3}. Thus, by using the transmon coupler, the static $ZZ$ coupling
below $20\,\rm{KHz}$ and the maintained $XY$ coupling strength above $2\,\rm{MHz}$
should be assessed experimentally for implementing cross-resonance gate with current fabrication
technology \cite{R37,R38}. We note that the maintained $XY$ coupling
may further increase by optimizing the full system parameters (see Appendix C for details).

\section{suppression of static ZZ interaction for on-resonantly coupled Transmons}\label{ch4}

\begin{figure}[tbp]
\begin{center}
\includegraphics[width=8cm,height=7.0cm]{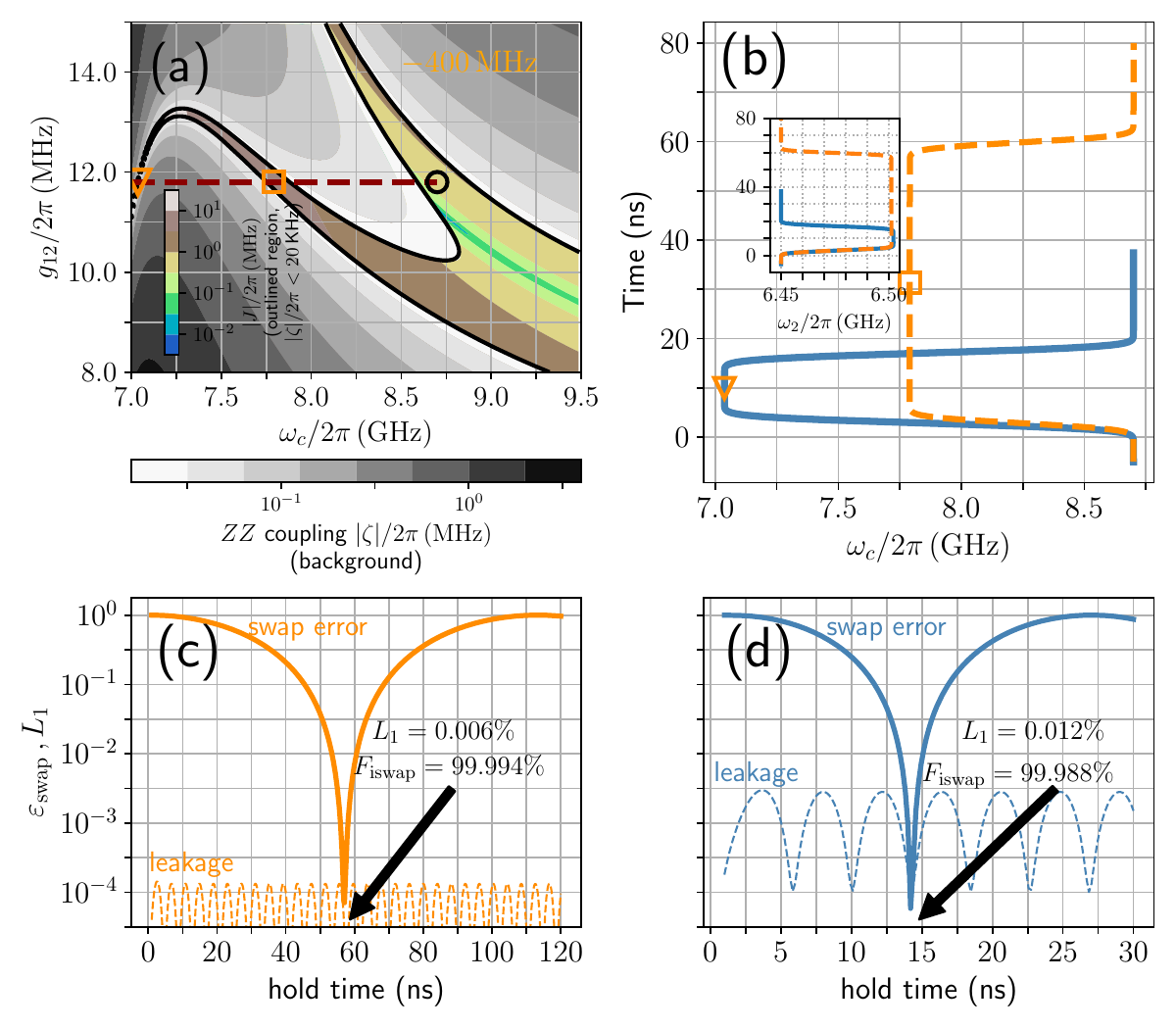}
\end{center}
\caption{ (a) Landscapes of numerically calculated $XY$ coupling strength $J$ and
$ZZ$ coupling strength $\zeta$ (date point with $ZZ$ coupling
strength below $20\,\rm{KHz}$ is omitted and replaced with the corresponding values of
the XY coupling $J$, i.e., the region outlined with black bold lines) as a function of coupler frequency $\omega_{c}$ and
direct coupling strength $g_{12}$ in the two-transmon system (Fig.~\ref{fig1}(b))
with qubit frequency $\omega_{1(2)}/2\pi=6.5\,\rm{GHz}$,
qubit anharmonicity $\alpha_{1(2)}/2\pi=-250\,\rm{MHz}$, coupler
anharmonicity $\alpha_{c}/2\pi={-400}\,\rm{MHz}$,
and frequency-dependent transmon-coupler coupling strength $g_{1c(2c)}/2\pi=125\sqrt{\omega_{c}/\omega_{1}}\,(130\sqrt{\omega_{c}/\omega_{2}})\,\rm{MHz}$ \cite{R39}.
The parameter set marked by an open circle denotes the idling frequency point for the coupler,
and the dashed line indicates the direct coupling strength adopted for mitigating the leakage during
the iSWAP gate operation. (b) Typical control pulse for implementing an iSWAP gate, where the full
width at half maximum is defined as hold time. The dashed (solid) line shows the control pulse for
system operated in dispersive (quasi-dispersive) regime, marked by the open square (triangle) in (a).
(c), (d) Swap error $\varepsilon_{\rm{swap}}$ and leakage $L_{1}$ as a function of hold time with
system operated in dispersive regime and quasi-dispersive regime, respectively.}
\label{fig4}
\end{figure}

Having shown the suppression of $ZZ$ coupling for the off-resonantly coupled case, we
now turn to consider the on-resonance case. Fig.~\ref{fig4}(a) shows the
numerically calculated $ZZ$ coupling strength $\zeta$ and $XY$ coupling strength $J$ (extracted as half the
energy difference between the eigenstate $|100\rangle$ and $|001\rangle$)
as a function of $\omega_{c}$ and $g_{12}$ for resonantly coupled transmons. Similar
to that of the off-resonantly coupled case, here, the zero-$ZZ$ points
also form two branches in the parameter space ($\omega_{c},\,g_{12}$) (see Appendix D for details),
where the lower branches allow us to mitigate $ZZ$ coupling for on-resonance $XY$
coupling based two-qubit gates.

To illustrate this, we consider the implementation of an iSWAP gate
with diabatic scheme \cite{R15,R39}, and in order to mitigate leakage,
here the direct coupling strength $g_{12}$ are chosen to achieve the full
synchronization between the swap and leakage error channels \cite{R39}. During the gate operations,
the frequency of the transmon $Q_{1}$ stays fixed
at $6.50\,\rm{GHz}$, while the frequencies of coupler $Q_{c}$ and transmon $Q_{2}$ vary
from their idle point ($\omega_{c(2)}/2\pi=8.70\,(6.45)\,\rm{GHz}$, where both
$XY$ and $ZZ$ coupling are strongly suppressed) to their interaction point and
then come back according to the Gaussian flat-top pulse with a fixed rise/fall time of 5.66 ns \cite{R40}.
We firstly consider that the system operates in the dispersive regime,
and coupler interaction frequency takes $\omega_{c}/2\pi\approx7.79\,\rm{GHz}$,
marked by the open square in Fig.~\ref{fig4}(a), and the associated control pulse is shown as the dashed
line in Fig.~\ref{fig4}(b). Fig.~\ref{fig4}(c) shows the swap error $\varepsilon_{\rm{swap}}=1-P_{001}$ ($P_{001}$
denotes the population in $|001\rangle$ after the time evolution for system initialized in $|100\rangle$) and the
leakage $L_{1}$ \cite{R41,R42} as function of the hold time that is defined as the time-interval between
the midpoints of the ramps \cite{R39}. One can find that an iSWAP gate with an intrinsic
gate fidelity \cite{R43} of $99.994\%$ and leakage $L_{1}$ below $0.006\%$ can be achieved with a hold
time of 57 ns. Moreover, by extracting conditional phase error $\delta_{\theta}$, we
find that $\delta_{\theta}$ is suppressed below $0.001$ rad, comforting strong suppression of
$ZZ$ coupling. In Fig.~\ref{fig4}(d), we also show the result
for the system operated in the quasi-dispersive regime \cite{R44,R45}, and the control pulse is shown
as the solid line in Fig.~\ref{fig4}(b), where $\omega_{c}/2\pi\approx7.04\,\rm{GHz}$ (also marked by the open circle
in Fig.~\ref{fig4}(a)), giving the $g_{1c(2c)}/\Delta_{1(2)}\approx1/4$. Although operating in the
quasi-dispersive regime, an iSWAP gate with fidelity above $99.988\%$ and leakage $L_{1}$
below $0.012\%$ can still be achieved with a hold time of 14.3 ns. The extracted conditional phase
error is suppressed below $0.003$ rad.

\section{conclusion}\label{ch5}

In summary, we have demonstrated that a feasible parameter region, where static
$ZZ$ coupling is heavily suppressed while leaving $XY$ interaction with an adequate strength
to implement two-qubit gates, such as cross-resonance gate or iSWAP gate, can be found in an
all-transmon system. We further show that an iSWAP gate with fast gate speed and dramatically
low conditional phase error can indeed be achieved in this parameter region. Without the detrimental
effect from the static $ZZ$ coupling, for transmon quantum processor with fixed coupling, single-qubit
addressing error, idling error, and crosstalk that arise from static $ZZ$ coupling should also be
heavily suppressed. From the point view of perturbation theory, the main physics behind these benefits
is that in the proposed system, $XY$ and $ZZ$ coupling are enabled by different virtual transitions
and different intermediate states, thus providing the possibility to engineer quantum
interference for mitigating $ZZ$ coupling while retaining $XY$ coupling. One thus reasonably
estimate that it is also possible to achieve the mitigation of static $ZZ$ coupling for $XY$-based
two-qubit gates with other types of coupler circuits \cite{R23} (see Appendix E for details).

Recently, two independent experimental works on suppression of $ZZ$ interactions for all-transmon qubit
systems have also been published \cite{R46,R47}. In the qubit architecture with tunable coupling,
Sung \emph{et al.} \cite{R46} have demonstrated experimentally that for qubit systems operated in the
quasi-dispersive regime (the qubit-coupler detuning should have a magnitude compared with that of
qubit anharmonicity), there is a working point (similar to the working point marked by a yellow
reversed triangle shown in Fig.~\ref{fig4}(a)), where the static $ZZ$ coupling is eliminated while $XY$
coupling is preserved, and based on the presence of this zero-$ZZ$ point, an $ZZ$-free iSWAP gate is
implemented with high gate fidelity. In this working point, the static $ZZ$ coupling is mainly resulted from
the interaction between qubit state $|101\rangle$ and its higher-energy state $|200(002)\rangle$
(giving a positive contribution) and coupler higher-energy state $|020\rangle$ (giving a negative contribution), thus when the positive
and negative contribution destructive interference, the static $ZZ$ interaction is eliminated. We note that
for qubit architectures with bus-mediated coupling operated in the quasi-dispersive straddling regime, there also
exists a zero-$ZZ$ point \cite{R44,R45}. However, as shown in Fig.~\ref{fig4}, in the qubit architecture
with tunable coupling, by choosing a suitable system parameters, there are three zero-$ZZ$ points, two of them
can be used to realize an $ZZ$-free iSWAP gate. For system comprising two off-resonantly coupled
transoms shown in Fig.~\ref{fig1}(b) (in \cite{R47}, the
coupler is a linear bus), Kandala \emph{et al.} \cite{R47} have shown
experimentally that when the qubit system operates in the dispersive regime, there exists a parameter
region (similar to the parameter region shown in Fig.~\ref{fig2}(d)), where one can
suppress the residual static $ZZ$ coupling while retain $XY$ coupling with an adequate strength for implementing
a successful CR gate. According to the analysis and numerical results given in Sec.~\ref{ch2} and Sec.~\ref{ch3},
we further show that by using a coupler with larger anharmonicity, i.e., replacing the linear bus with an
ancillary transmon, the retained $XY$ coupling should have a more larger strength. The physics behind how
this works is that when using a transmon coupler, the energy of $|020\rangle$ is reduced, and thus the
negative $ZZ$ contribution ($\zeta_{020}$) from interaction $|101\rangle\leftrightarrow|020\rangle$ gets
larger. Therefore, in this case, a larger retained $XY$ coupling $J$ is needed to eliminate $ZZ$ coupling.

\begin{acknowledgments}
We thank Ji Chu for insightful discussion.
This work was partly supported by the National Key Research and Development Program of
China (Grant No.2016YFA0301802), the National Natural Science Foundation of
China (Grant No.61521001, and No.11890704), and the Key R$\&$D Program of
Guangdong Province (Grant No.2018B030326001). X. T. acknowledges the supported by
the National Natural Science Foundation of
China (Grant No.12074179). P. X. acknowledges the supported by
the Young fund of Jiangsu Natural Science
Foundation of China (Grant No.BK20180750). H. Y. acknowledges support from the
Beijing Natural Science Foundation (Grant No.Z190012).

P. Z. and D. L. contributed equally to this work.
\end{acknowledgments}

\appendix

\section{strength of static $ZZ$ coupling}

\begin{figure}[tbp]
\begin{center}
\includegraphics[width=8cm,height=3cm]{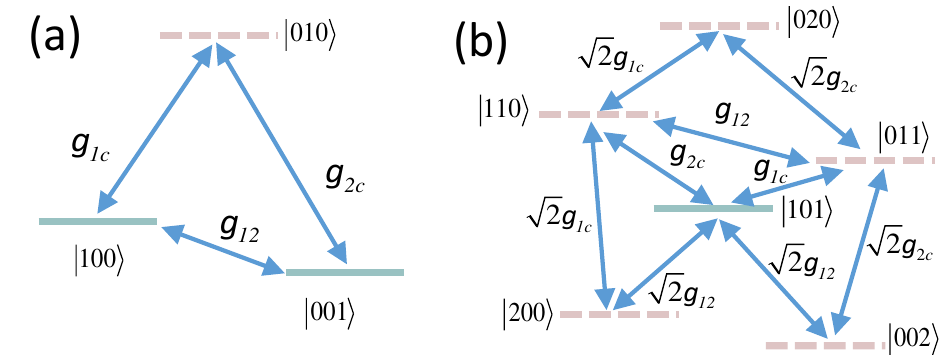}
\end{center}
\caption{Level diagram of the coupled transmon system. (a) one-excitation subspace.
(b) two-excitation subspace. The green solid lines denote qubit level, while the pink
dashed lines represent the non-qubit levels.}
\label{fig5}
\end{figure}

In this section, according to perturbation theory \cite{R32}, we present the details deviation of the static $ZZ$
coupling strength $\zeta$ for our two-transmon system. In order to give a clear explanation of how to
mitigating $ZZ$ coupling for $XY$-based gate operations, in the following discussion, we have made
two approximations: (i) Making the rotating-wave approximation (RWA) by neglecting
the counter-rotating terms in Eq.$\,$(1) of the main text, the system Hamiltonian now reads $H=H_{0}+V$, with
\begin{eqnarray}
\begin{aligned}
\label{eq4}
&H_{0}=\sum_{j}\left[\tilde{\omega}_{j}q_{j}^{\dagger}q_{j}+\frac{\alpha_{j}}{2}q_{j}^{\dagger}q_{j}^{\dagger}q_{j}q_{j}\right],
\\&V=\sum_{j,k}g_{jk}(q_{j}^{\dagger}q_{k}+q_{j}q_{k}^{\dagger}),
\end{aligned}
\end{eqnarray}
and its level diagram is shown in Fig.~\ref{fig5}.
(ii) Since $g_{12}\ll\{g_{1c},g_{2c}\}$ and $\Delta_{1(2)}\gg \{\Delta_{12},\alpha_{1,2,c}\}$, one can
neglect small terms in the calculation of $ZZ$ coupling strength $\zeta$.

The perturbed result for $ZZ$ coupling strength $\zeta$ can be defined as $\zeta\equiv\zeta^{(2)}+\zeta^{(3)}+\zeta^{(4)}$,
where $\zeta^{(n)}$ denotes $n$th-order perturbational result, defined as $\zeta^{(n)}\equiv(E^{(n)}_{101}-E^{(n)}_{001})-(E^{(n)}_{100}-E^{(n)}_{000})$
with
\begin{eqnarray}
\begin{aligned}
\label{eq5}
&E^{(2)}_{s}=\sum_{j\neq s}\frac{|V_{sj}|^{2}}{E_{sj}},
\\&E^{(3)}_{s}=\sum_{j,k\neq s}\frac{V_{sj}V_{jk}V_{ks}}{E_{sj}E_{sk}},
\\&E^{(4)}_{s}=\sum_{j,k,l\neq s}\frac{V_{sj}V_{jk}V_{kl}V_{ls}}{E_{sj}E_{sk}E_{sl}}
-\sum_{j,k\neq s}\frac{|V_{sj}|^{2}|V_{sk}|^{2}}{E_{sj}^{2}E_{sk}},
\end{aligned}
\end{eqnarray}
where $V_{sj}=\langle s|V|j\rangle$ and $E_{sj}=E^{(0)}_{s}-E^{(0)}_{j}$. Thus, after making the
above mentioned two approximations, and according to the expression in Eq.~(\ref{eq5}), one has \cite{R32,R33,R34}
\begin{equation}
\begin{aligned}
\label{eq6}
\zeta^{(2)}=2g_{12}^{2}\big[\frac{1}{\Delta_{12}-\alpha_{2}}-\frac{1}{\Delta_{12}+\alpha_{1}}\big],
\end{aligned}
\end{equation}

\begin{equation}
\begin{aligned}
\label{eq7}
\zeta^{(3)}=2g_{12}g_{1c}g_{2c}\big[&\frac{2}{(\Delta_{12}-\alpha_{2})\Delta_{1}}-\frac{2}{(\Delta_{12}+\alpha_{1})\Delta_{2}}
\\&+\frac{2}{\Delta_{1}\Delta_{2}}\big],
\end{aligned}
\end{equation}

\begin{eqnarray}
\begin{aligned}
\label{eq8}
\zeta^{(4)}=2g_{1c}^{2}g_{2c}^{2}\big[&\frac{1}{\Delta_{1}^{2}(\Delta_{12}-\alpha_{2})}-\frac{1}{\Delta_{2}^{2}(\Delta_{12}+\alpha_{1})}
\\&+\frac{1}{\Delta_{1}+\Delta_{2}-\alpha_{c}}(\frac{1}{\Delta_{1}}+\frac{1}{\Delta_{2}})^{2}\big].
\end{aligned}
\end{eqnarray}

To identify the physical mechanism behind there terms ($\zeta^{(2)},\zeta^{(3)},\zeta^{(4)}$), and also to give a clear
analysis of the relation between $ZZ$ coupling strength $\zeta$ and $XY$ coupling strength $J$,
after writing out all these terms and rearranging them, $\zeta$ can be approximated as (here we recover the
Eq.~(\ref{eq3}) of the main text)
\begin{eqnarray}
\begin{aligned}
\label{eq9}
&\zeta\simeq \zeta_{020}+\zeta_{200}+\zeta_{002}+\zeta_{1},
\\&\zeta_{020}=\frac{J_{020}^{2}}{\Delta_{1}+\Delta_{2}-\alpha_{c}},\zeta_{200}=\frac{J_{200}^{2}}{\Delta_{12}-\alpha_{2}},
\\&\zeta_{002}=-\frac{J_{002}^{2}}{\Delta_{12}+\alpha_{1}},\zeta_{1}=\frac{4g_{12}g_{1c}g_{2c}}{\Delta_{1}\Delta_{2}},
\end{aligned}
\end{eqnarray}
where terms $\zeta_{020}$ $\zeta_{002(200)}$ can be considered as the $ZZ$ contributions resulting from the effective
coupling between qubit state $|101\rangle$ and higher-energy states of coupler $|020\rangle$ and transmons $|002(200)\rangle$,
respectively, and $J_{020},\,J_{200(002)}$ denote the associated effective coupling strength, given as
\begin{eqnarray}
\begin{aligned}
\label{eq10}
&J_{020}\simeq \sqrt{2}g_{1c}g_{2c}\big(\frac{1}{\Delta_{1}}+\frac{1}{\Delta_{2}}\big)\simeq\frac{2\sqrt{2}g_{1c}g_{2c}}{\Delta},
\\&J_{200}\simeq \sqrt{2}\big(g_{12}+\frac{g_{1c}g_{2c}}{\Delta_{1}}\big)\simeq\sqrt{2}J,
\\&J_{002}\simeq \sqrt{2}\big(g_{12}+\frac{g_{1c}g_{2c}}{\Delta_{2}}\big)\simeq\sqrt{2}J,
\end{aligned}
\end{eqnarray}
while term $\zeta_{1}$ results from the interaction among lower-energy states of qubits and coupler.

\section{Estimated XY coupling strength from cross-resonance oscillation}
\begin{figure}[tbp]
\begin{center}
\includegraphics[width=8cm,height=3.5cm]{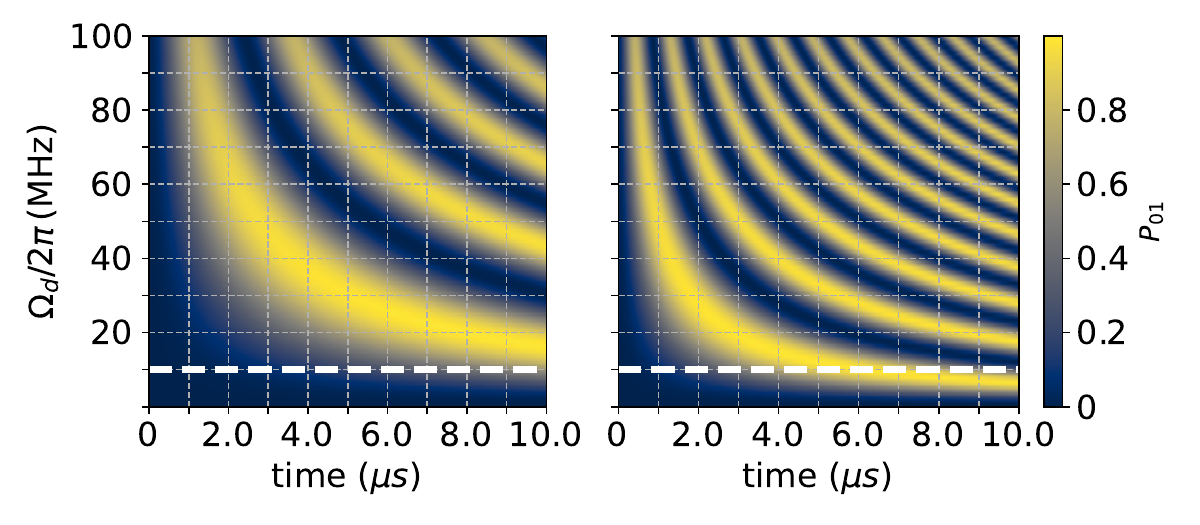}
\end{center}
\caption{Cross-resonance oscillation of target transmon $Q_{2}$ with the controlled
transmon $Q_{1}$ in its ground state. (Left/Right panel) Population oscillation for
target transmon $Q_{2}$ as a function of strength $\Omega_{d}$ of the driving applied on transmon $Q_{1}$
and time with parameter set $(\omega_{c},\,g_{12})$ marked with open circle/square
shown in Figs.~\ref{fig2}(c,d) of the main text. The other system parameters are similar to those used in
Fig.~\ref{fig2} of the main text. The horizontal dashed lines indicates the driving strength
$\Omega_{d}/2\pi=10\,\rm{MHz}$ that is used to infer the effective XY coupling strength $J$ \cite{R36}.}
\label{fig6}
\end{figure}

For two on-resonantly coupled qubits, the inter-qubit $XY$ coupling strength can be extracted
as half the energy difference between the eigenstate $|100\rangle$ and $|001\rangle$.
However, for two off-resonant qubits coupled via a coupler circuit shown in Fig~\ref{fig1}(b) of
the main text, the effective $XY$ coupling is not perfectly well-defined in this case. In present
work, the $XY$ coupling strength is estimated from the period $T$ of the cross-resonance
oscillation with the controlled qubit $Q_{1}$ in its ground state, as shown in Fig.~\ref{fig6}. In the
weak-drive limit, the period $T$ of the oscillations can be well approximated
by $2\pi/T=J\Omega_{d}/\Delta_{12}$ \cite{R36}, where $\Omega_{d}$ denotes strength of the driving
applied on $Q_{1}$. In present work, $\Omega/2\pi=10\,\rm{MHz}$ is used to infer the $XY$
coupling strength.

\section{off-resonantly coupled qubits with varying qubit detuning}

\begin{figure*}[tbp]
\begin{center}
\includegraphics[width=18cm,height=8cm]{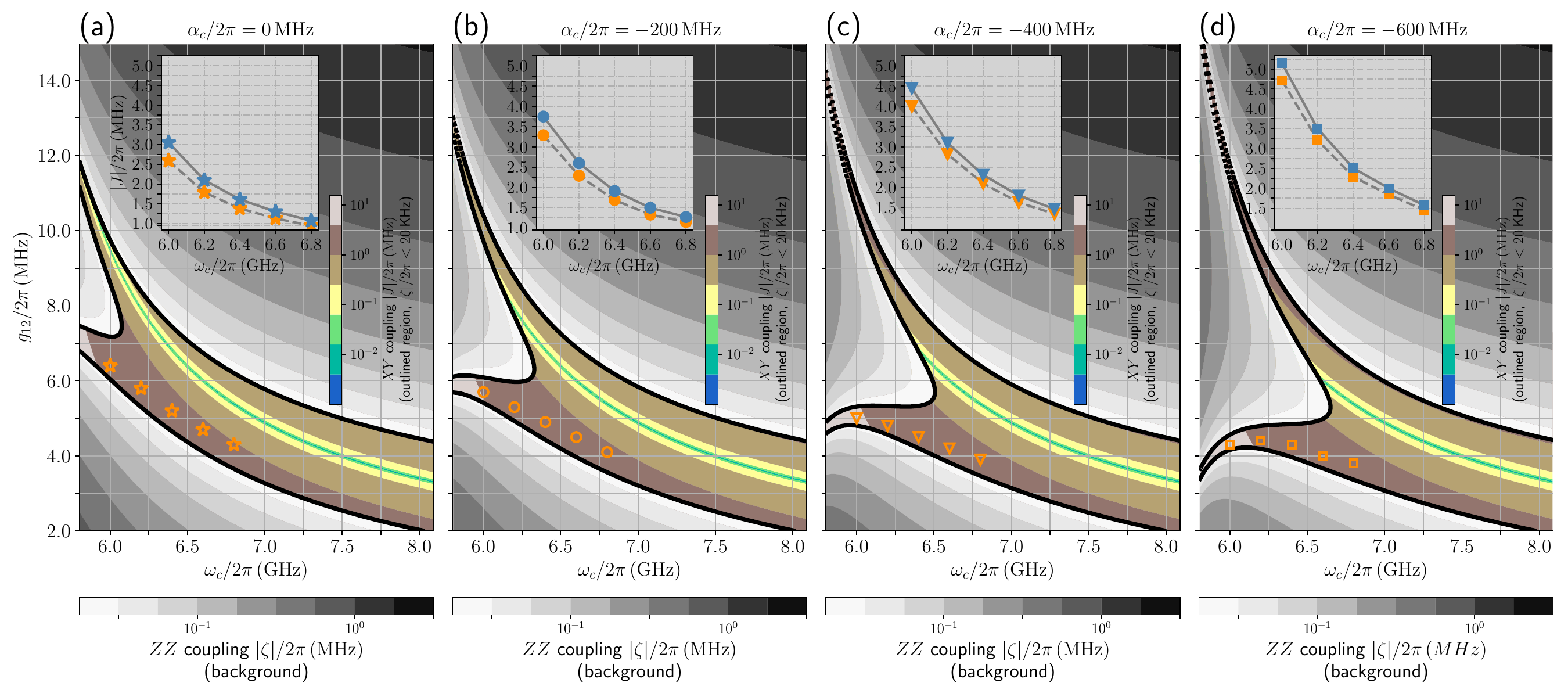}
\end{center}
\caption{Landscapes of $XY$ coupling $J$ and $ZZ$ coupling $\zeta$
 as a function of coupler frequency $\omega_{c}$ and direct coupling strength $g_{12}$ with
qubit frequency $\omega_{1(2)}/2\pi=5.114\,(5.014)\,\rm{GHz}$. The other system parameters are similar to
those used in Fig.~\ref{fig3} of the main text, where the qubit detuning $\Delta_{12}/2\pi=200\,\rm{MHz}$.
Here the two-qubit detuning $\Delta_{12}/2\pi=100\,\rm{MHz}$.}
\label{fig7}
\end{figure*}

\begin{figure*}[tbp]
\begin{center}
\includegraphics[width=18cm,height=8cm]{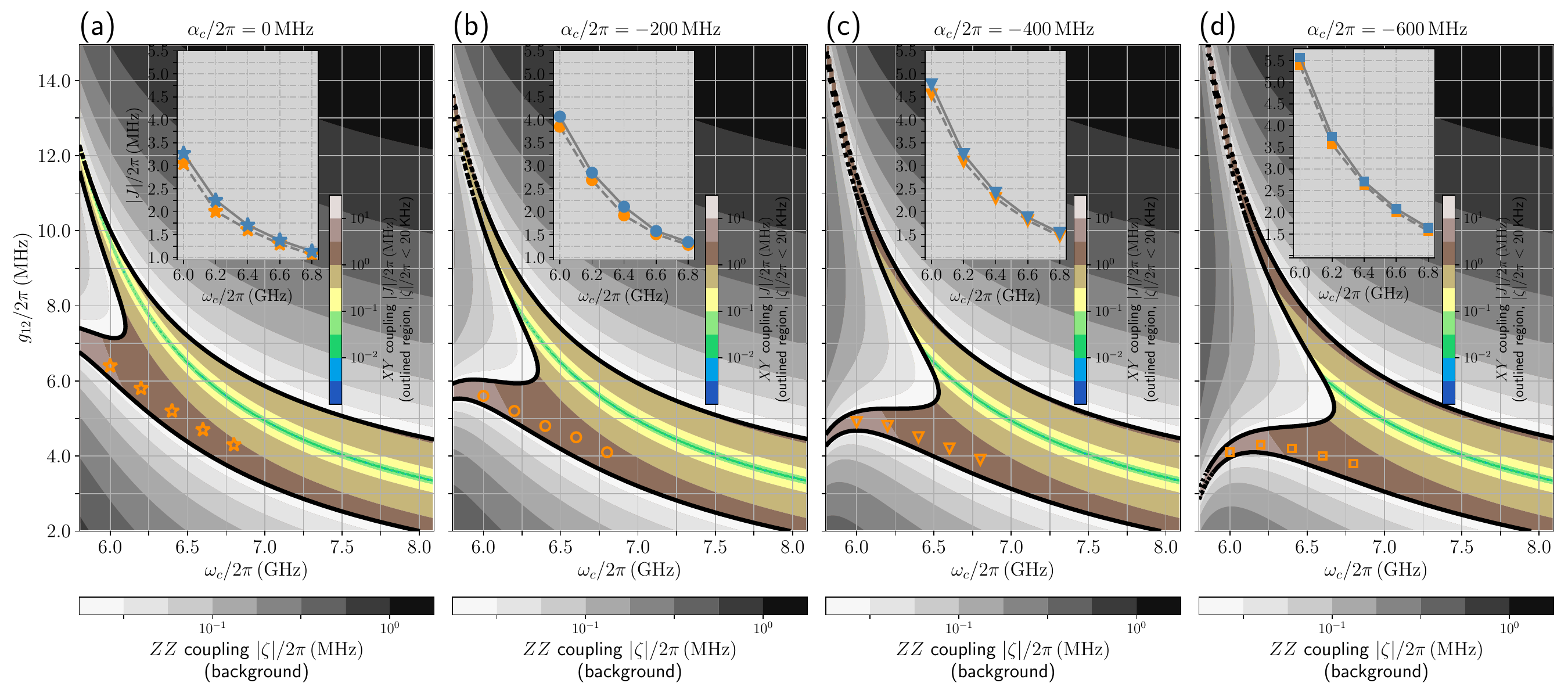}
\end{center}
\caption{Landscapes of $XY$ coupling $J$ and $ZZ$ coupling $\zeta$
 as a function of coupler frequency $\omega_{c}$ and direct coupling strength $g_{12}$ with
qubit frequency $\omega_{1(2)}/2\pi=5.114\,(5.064)\,\rm{GHz}$. The other system parameters are similar to
those used in Fig.~\ref{fig3} of the main text, where the qubit detuning $\Delta_{12}/2\pi=200\,\rm{MHz}$.
Here the two-qubit detuning $\Delta_{12}/2\pi=50\,\rm{MHz}$. }
\label{fig8}
\end{figure*}

As we have mentioned in the main text, the strength of the maintained $XY$ coupling may further increase by
optimizing the full system parameters. In Figs.~\ref{fig7} and~\ref{fig8}, we have shown the landscapes of $XY$ coupling $J$
and $ZZ$ coupling $\zeta$ as a function of coupler frequency $\omega_{c}$ and direct coupling strength $g_{12}$
for off-resonantly coupled qubits system with different qubit detuning,i.e., $\Delta_{12}/2\pi=100\,\rm{MHz}$,
and $\Delta_{12}/2\pi=50\,\rm{MHz}$, respectively.

\section{on-resonantly coupled qubits with varying coupler anhrmonicity }

\begin{figure*}[tbp]
\begin{center}
\includegraphics[width=18cm,height=8cm]{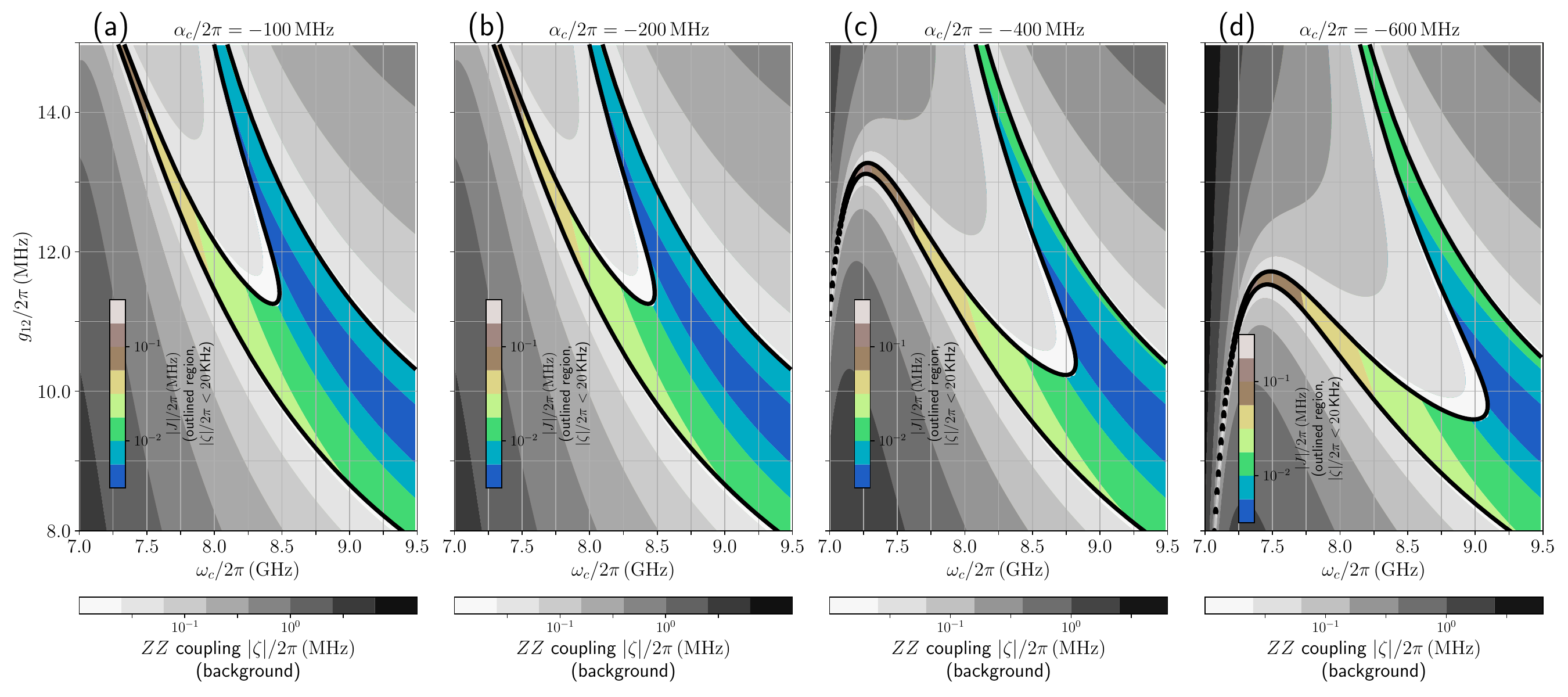}
\end{center}
\caption{Landscapes of $XY$ coupling $J$ and $ZZ$ coupling $\zeta$
as a function of coupler frequency $\omega_{c}$ and direct coupling strength $g_{12}$ for on-resonantly
coupled qubits. (a) $\alpha_{c}/2\pi=-100\,\rm{MHz}$, (b) $\alpha_{c}/2\pi=-200\,\rm{MHz}$,
(c) $\alpha_{c}/2\pi=-400\,\rm{MHz}$ (see also in Fig.~\ref{fig4}(a) of the main text), and (d)
$\alpha_{c}/2\pi=-600\,\rm{MHz}$. The other system parameters are similar to those used in Fig.~\ref{fig4}(a)
of the main text. }
\label{fig9}
\end{figure*}

Similar to the off-resonantly coupled case, when we reduce the energy of $|020\rangle$, i.e.,
increasing the coupler anharmonicity for on-resonantly coupled transmon system, the
negative $ZZ$ contribution ($\zeta_{020}$) from interaction $|101\rangle\leftrightarrow|020\rangle$
gets larger, thus a larger maintained $XY$ coupling $J$ is needed to suppress $ZZ$ coupling.
Figure~\ref{fig9} show the numerical calculated $ZZ$ coupling strength and the $XY$ coupling
strength as a function of $\omega_{c}$ and $g_{12}$ with coupler anharmonicity
$\alpha_{c}/2\pi=\{0,\,-200,\,-400,\,-600\}\,\rm{MHz}$. The other parameters are
same to those used in Fig.~\ref{fig4}(a) of the main text. One can indeed find
that the lower branch of zero-$ZZ$ region is shifted downward along with the increased
coupler anharmonicity, suggesting a larger maintained $XY$ coupling.

\section{Tunable coupling superconducting circuit}

In this section, we show that suppressing $ZZ$ coupling while preserving $XY$ coupling can also be realized
in the qubit architecture proposed by P. S. Mundada \emph{et al.} \cite{R23}, where two transoms are
coupled via a coupling circuit comprising two bus couplers. According to Ref.$\,$\cite{R23}, the full
system can be modeled as four coupled weakly anharmonic oscillators, and its Hamiltonian reds
\begin{equation}
\begin{aligned}
\label{eq11}
H=&\sum_{j=1,2,\pm}\big(\tilde{\omega}_{j}q_{j}^{\dagger}q_{j}+\frac{\alpha_{j}}{2}q_{j}^{\dagger}q_{j}^{\dagger}q_{j}q_{j}\big)
\\&+\sum_{\substack{j=1,2\\k=\pm}}g_{jk}(q_{j}q_{k}^{\dagger}+q_{j}^{\dagger}q_{k}),
\end{aligned}
\end{equation}
where subscript $j(k)=\{1,2,\pm\}$ labels anharmonic oscillator $Q_{j}$ with anharmonicity
$\alpha_{j}$ and bare transition frequency $\tilde{\omega}_{j}$, $q_{j}\,(q_{j}^{\dagger})$ is
the associated annihilation (creation) operator, and $g_{jk}$ denotes
strength of the coupling between $Q_{j}$ and $Q_{k}$.

From second-order perturbation theory, the $XY$ coupling strength can be obtained as
\begin{equation}
\begin{aligned}
\label{eq12}
&J=J_{+}+J_{-},
\\&J_{\pm}=g_{1\pm}g_{2\pm}/\Delta_{\pm},
\end{aligned}
\end{equation}
with $1/\Delta_{\pm}=(1/\Delta_{1\pm}+1/\Delta_{2\pm})/2$.
The $ZZ$ coupling can be defined as $\zeta = (E_{1100} - E_{1000}) - (E_{0100}-E_{0000})$, where $E_{n_{1}n_{2}n_{-}n_{+}}$
denotes the energy of system eigenstate $|n_{1}n_{2}n_{-}n_{+}\rangle$ ($n_{1},n_{2},n_{-},n_{+}=\{0,1,2\}$).
According to the fourth order perturbation theory \cite{R23,R32}, the expression for $\zeta$
is $\zeta=\zeta_{2000} + \zeta_{0200}+\zeta_{0020}+\zeta_{0002}+\zeta_{1}$ with
\begin{widetext}
\begin{equation}
\begin{aligned}
\label{eq13}
&\zeta_{2000}=\left(\frac{g_{1+}g_{2+}}{\Delta_{2+}}+\frac{g_{1-}g_{2-}}{\Delta_{2-}}\right)^2\left(\frac{2}{-\Delta_{12}-\alpha_1}\right)
=\frac{J^{2}_{2000}}{-\Delta_{12}-\alpha_1},
\\&\zeta_{0200}=\left(\frac{g_{1+}g_{2+}}{\Delta_{1+}}+\frac{g_{1-}g_{2-}}{\Delta_{1-}}\right)^2\left(\frac{2}{\Delta_{12}-\alpha_2}\right)
=\frac{J^{2}_{0200}}{\Delta_{12}-\alpha_2},
\\&\zeta_{0020}=\frac{2 g_{1+}^2 g_{2+}^2}{\Delta_{1+}+\Delta_{2+}+\alpha_+}\left(\frac{1}{\Delta_{1+}}+\frac{1}{\Delta_{2+}}\right)^2
=\frac{J^{2}_{0020}}{\Delta_{1+}+\Delta_{2+}+\alpha_+},
\\&\zeta_{0002}=\frac{2 g_{1-}^2 g_{2-}^2}{\Delta_{1-}+\Delta_{2-}+\alpha_-}\left(\frac{1}{\Delta_{1-}}+\frac{1}{\Delta_{2-}}\right)^2
=\frac{J^{2}_{0002}}{\Delta_{1-}+\Delta_{2-}+\alpha_-},
\\&\zeta_{1}=\frac{1}{\Delta_{12}}\left[\left(\frac{g_{1+}g_{2+}}{\Delta_{1+}}+\frac{g_{1-}g_{2-}}{\Delta_{1-}}\right)^2-
\left(\frac{g_{1+}g_{2+}}{\Delta_{2+}}+\frac{g_{1-}g_{2-}}{\Delta_{2-}}\right)^2\right]
\\&\quad+\left[g_{1+}g_{2-}\left(\frac{1}{\Delta_{1+}}+\frac{1}{\Delta_{2-}}\right)+g_{1-}g_{2+}\left(\frac{1}{\Delta_{1-}}+\frac{1}{\Delta_{2+}}\right)\right]^2\frac{1}{\Delta_{1+}+\Delta_{2-}}\\
&\quad - \left(\frac{g_{1+}^2}{\Delta_{1+}^2}+\frac{g_{1-}^2}{\Delta_{1-}^2}\right)\left(\frac{g_{2+}^2}{\Delta_{2+}}+\frac{g_{2-}^2}{\Delta_{2-}}\right)
-\left(\frac{g_{2+}^2}{\Delta_{2+}^2}+\frac{g_{2-}^2}{\Delta_{2-}^2}\right)\left(\frac{g_{1+}^2}{\Delta_{1+}}+\frac{g_{1-}^2}{\Delta_{1-}}\right),
\end{aligned}
\end{equation}
\end{widetext}
where $\Delta_{ij}=\tilde{\omega}_{i}-\tilde{\omega}_{j}$. The terms $\zeta_{2000(0200)}$ and $\zeta_{0020(0002)}$ can be considered as the $ZZ$ contributions resulting from the effective
coupling between qubit state $|1100\rangle$ and higher-energy states of tranmsons $|2000(0200)\rangle$ and coupled $|0020(0002)\rangle$,
respectively, and the associated effective coupling strength can be approximated as
\begin{eqnarray}
\begin{aligned}
\label{eq14}
&J_{2000}\simeq \sqrt{2}J,J_{0200}\simeq \sqrt{2}J,
\\&J_{0020}\simeq 2\sqrt{2}J_{+},J_{0002}\simeq 2\sqrt{2}J_{-},
\end{aligned}
\end{eqnarray}
while term $\zeta_{1}$ results from the interaction among lower-energy states of qubits and coupler.

\begin{figure*}[tbp]
\begin{center}
\includegraphics[width=18cm,height=7cm]{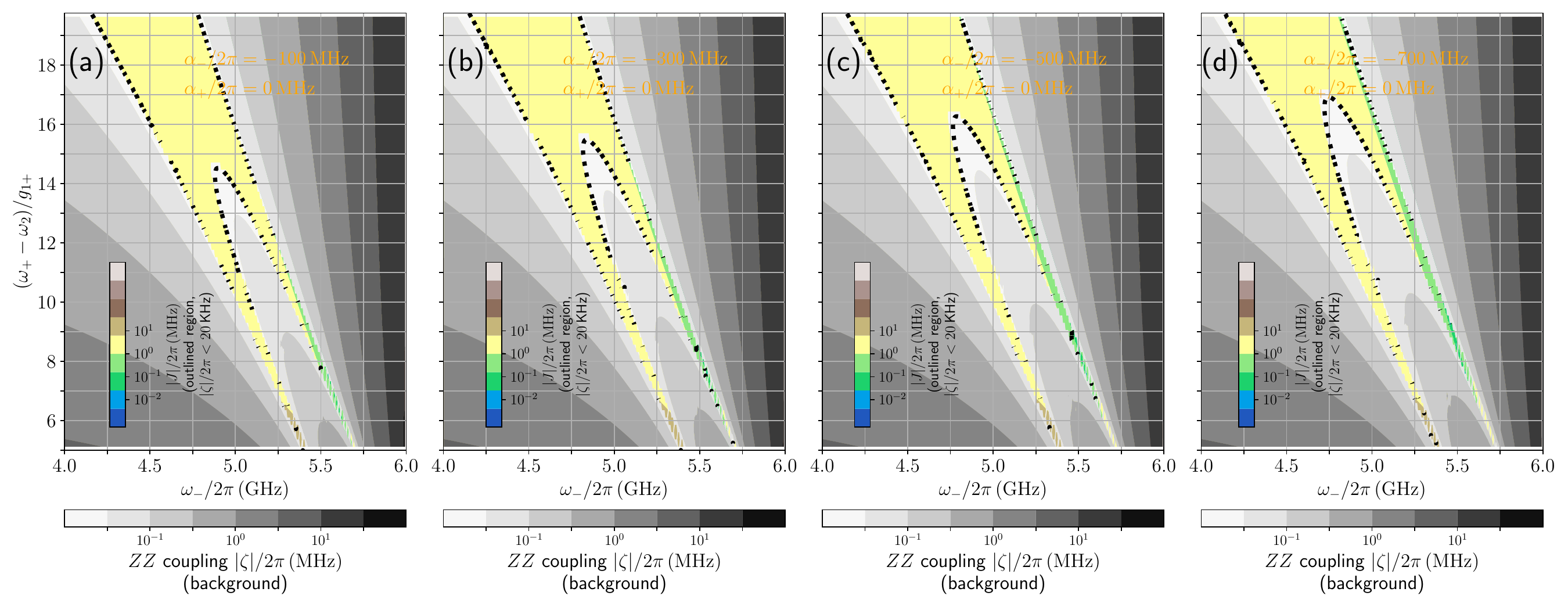}
\end{center}
\caption{Off-resonance case. Landscapes of $XY$ coupling $J$ (perturbation theory) and $ZZ$ coupling $\zeta$
(numerical diagonalization) as a function of coupler frequency $\omega_{\pm}$ with
qubit frequency $\omega_{1(2)}/2\pi=6.143\,(6.421)\,\rm{GHz}$,
qubit anharmonicity $\alpha_{1(2)}/2\pi=-330\,\rm{MHz}$, coupler
anharmonicity $\alpha_{+}/2\pi=0\,\rm{MHz}$, transmon-coupler
coupling strength $g_{1-(2-)}/2\pi=85\,\rm{MHz}$, $g_{1+(2+)}/2\pi=102\,\rm{MHz}$ \cite{R5} and coupler anharmonicity
(a) $\alpha_{-}/2\pi=-100\,\rm{MHz}$, (b) $\alpha_{c}/2\pi=-300\,\rm{MHz}$,
(c) $\alpha_{c}/2\pi=-500\,\rm{MHz}$, and (d) $\alpha_{c}/2\pi=-700\,\rm{MHz}$. For each coupler anharmonicity, the two landscapes
are superposed in a single panel, where the region corresponding to ZZ coupling with $\zeta$ below 20 KHz is
omitted and replaced with the corresponding values of the XY coupling $J$ (the region outlined with black dotted lines).}
\label{fig10}
\end{figure*}

\begin{figure*}[tbp]
\begin{center}
\includegraphics[width=18cm,height=7cm]{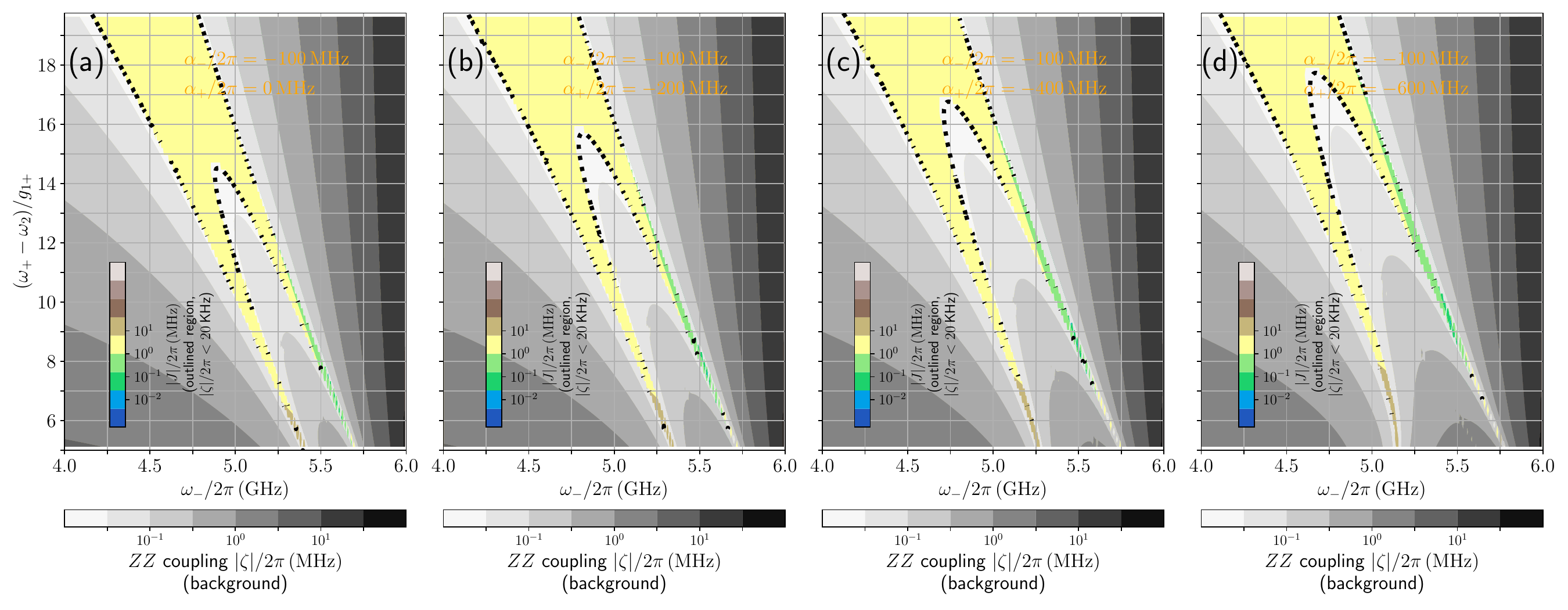}
\end{center}
\caption{Off-resonance case. Landscapes of $XY$ coupling $J$ (perturbation theory) and $ZZ$
coupling $\zeta$ (numerical diagonalization) as a function of coupler frequency $\omega_{\pm}$ with coupler
anharmonicity $\alpha_{-}/2\pi=-100\,\rm{MHz}$. (a) $\alpha_{+}/2\pi=0$, (b) $\alpha_{+}/2\pi=-200\,\rm{MHz}$,
(c) $\alpha_{+}/2\pi=-400\,\rm{MHz}$, and (d) $\alpha_{+}/2\pi=-600\,\rm{MHz}$. The other system parameters are
similar to those used in Fig.~\ref{fig10}.}
\label{fig11}
\end{figure*}

\begin{figure*}[tbp]
\begin{center}
\includegraphics[width=18cm,height=7cm]{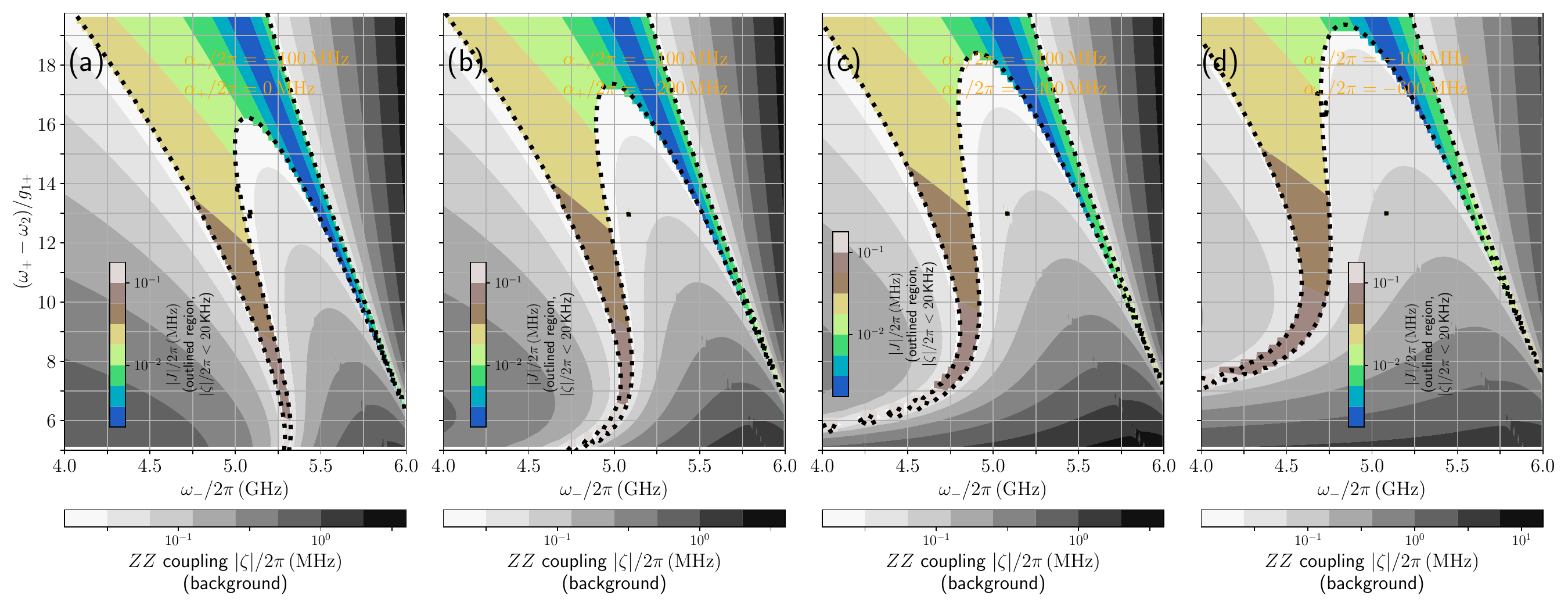}
\end{center}
\caption{On-resonance case. Landscapes of $XY$ coupling $J$ (numerical diagonalization) and $ZZ$
coupling $\zeta$ (numerical diagonalization) as a function of coupler frequency $\omega_{\pm}$ with
qubit frequency $\omega_{1(2)}/2\pi=6.421\,\rm{GHz}$. The other system parameters are
similar to those used in Fig.~\ref{fig11}.}
\label{fig12}
\end{figure*}

From Eqs.~(\ref{eq12}),~(\ref{eq13}) and~(\ref{eq14}), one can find that the conditions for achieving zero-$XY$ coupling and
zero-$ZZ$ coupling is not coexist in the same parameter space. This suggests that in this coupler
architecture, one may suppress static $ZZ$ coupling while preserve $XY$ interaction.
In Figs.~\ref{fig10},~\ref{fig11}, and~\ref{fig12}, we demonstrate numerically that in this architecture, the staic $ZZ$
coupling can indeed be heavily suppressed without the need for suppressing $XY$ coupling. Thus,
$XY$-based two-qubit gates could be implemented without the detrimental effect from static $ZZ$
coupling.

\begin{thebibliography}{99}
\bibitem{R31} In present work, our discussion is restricted on all-transmon system where qubits are in the straddling regime.
    For transmons out of straddling regime, the previous numerical result does not show any possibility for achieving suppression of static $ZZ$ coupling in all-transmon device operated in a reasonable coupling regime \cite{R23}.
\end{thebibliography}
\end{document}